\documentclass[twocolumn,superscriptaddress,amsmath,amssymb,aps,prl]{revtex4-2}
\usepackage{amsmath,amssymb}
\usepackage{graphicx}
\usepackage{dcolumn}
\usepackage{bm}
\usepackage{multirow}
\usepackage{textcomp}
\usepackage{url}
\usepackage{enumitem}
\usepackage{float}
\usepackage[dvipsnames]{xcolor}
\usepackage[colorlinks=true, allcolors=blue!80!black!]{hyperref}
\usepackage{orcidlink}

\newcommand{\ket}[1]{|{#1}\rangle}

\begin{document}

\title{Full Configuration Interaction Quantum Monte Carlo for Accurate \textit{Ab Initio} Nuclear Structure Calculations}

\author{R. Z. Hu\,\orcidlink{0009-0002-8797-6622}}
\affiliation{School of Physics, and State Key Laboratory of Nuclear Physics and Technology, Peking University, Beijing 100871, China}
\author{F. R. Xu\,\orcidlink{0000-0001-6699-0965}}\email[]{frxu@pku.edu.cn}
\affiliation{School of Physics, and State Key Laboratory of Nuclear Physics and Technology, Peking University, Beijing 100871, China}
\affiliation{Southern Center for Nuclear-Science Theory (SCNT), Institute of Modern Physics, Chinese Academy of Sciences, Huizhou 516000, China}
\author{B. S. Hu\,\orcidlink{0000-0001-8071-158X}}\email[]{baishanhu@pku.edu.cn}
\affiliation{School of Physics, and State Key Laboratory of Nuclear Physics and Technology, Peking University, Beijing 100871, China}
\affiliation{Southern Center for Nuclear-Science Theory (SCNT), Institute of Modern Physics, Chinese Academy of Sciences, Huizhou 516000, China}
\author{A. Alavi\,\orcidlink{0000-0002-0654-9489}}
\affiliation{Max Planck Institute for Solid State Research, Heisenbergstr. 1, 70569 Stuttgart, Germany}
\affiliation{Department of Chemistry, University of Cambridge, Lensfield Road, Cambridge CB2 1EW, United Kingdom}

\date{\today}

\begin{abstract}
    We introduce novel full configuration interaction quantum Monte Carlo (FCIQMC) as an accurate many-body solver for \textit{ab initio} nuclear structure calculations. This stochastic approach directly samples the exact wave function in the full configuration space, enabling high-fidelity treatment of high-order many-body correlations in strongly interacting nuclear systems. Using interactions from chiral effective field theory, we have computed ground-state energies and charge radii of $^4$He, $^8$Be, $^{12}$C and $^{16}$O with sub-percent-level many-body uncertainties. These results establish FCIQMC as a stochastic full-configuration-space solver capable of treating systems beyond the reach of the conventional no-core shell model, and as an accurate benchmark for truncated many-body expansion methods.
\end{abstract}

\maketitle

\textit{Introduction.\textemdash}
The development and application of \textit{ab initio} many-body methods for strongly interacting atomic nuclei are among the most exciting areas in modern nuclear theory~\cite{Hergert2020}. Within the framework of chiral effective field theory~\cite{Epelbaum2009,Machleidt2011,Hammer2020,Tews2022}, nuclear interactions and currents~\cite{PhysRevC.78.064002,PhysRevLett.103.102502,PhysRevC.80.034004,PhysRevLett.107.062501,Gysbers2019,PhysRevC.102.025501,Krebs2020,PhysRevLett.132.232503,PhysRevLett.132.232504} can be derived in a systematic way from the fundamental theory of strong interactions, quantum chromodynamics. The scope of \textit{ab initio} calculations has been extended from light nuclei to medium-mass nuclei and even to several heavy-mass nuclei over the past decade~\cite{Binder2014,PhysRevLett.120.152503,PhysRevLett.125.182501,Gysbers2019,PhysRevLett.126.042502,Hu2022np,PhysRevC.105.014302,PhysRevC.107.024310,Karthein2024,PhysRevLett.134.063002,bonaiti2025,Jiang2026}.

The no-core shell model (NCSM)~\cite{Barrett2013,PhysRevLett.84.5728,PhysRevC.79.014308} is one of the earliest successful many-body methods from first principles, which solves the large-scale eigenvalue problem in the truncated configuration space. However, the NCSM is severely limited by the rapidly growing model-space dimension, which easily reaches the computational limit of current computers of $\sim 10^{12}$~\cite{Maris2010}. As a result, direct NCSM computations are mostly restricted to light nuclei, and rely on extrapolations with respect to the model-space size~\cite{PhysRevC.91.061301,PhysRevC.105.L061302,6zk6-1sy6}. Several techniques have been developed to extend the reach of the NCSM, including the use of natural orbitals~\cite{PhysRevC.99.034321}, importance truncation scheme~\cite{PhysRevLett.99.092501,PhysRevC.87.044301,PhysRevC.93.021301}, symmetry-adapted scheme~\cite{PhysRevLett.98.162503,LAUNEY2016101,PhysRevLett.125.102505,c3st-tp13} and neural network extrapolations~\cite{PhysRevC.99.054308,PhysRevC.110.014327,PhysRevC.111.064304,8mfb-wc36}.

There is another family of many-body methods that is based on systematic many-body expansions on top of a reference state, including many-body perturbation theory (MBPT)~\cite{ROTH2010272,PhysRevC.86.054315,PhysRevC.94.014303,TICHAI2018448,PhysRevLett.122.042501,PhysRevC.95.034326,10.3389/fphy.2020.00164}, in-medium similarity renormalization group (IMSRG)~\cite{PhysRevLett.106.222502,Hergert2016,stroberg2017,PhysRevC.105.L061303,PhysRevC.99.061302,Zhen2025}, coupled-cluster theory (CC)~\cite{kuemmel1978,RevModPhys.79.291,shavittbartlett2009,Hagen2014,PhysRevC.76.034302,PhysRevLett.113.142502,sun2025,p297-y8vq} and self-consistent Green's function (SCGF)~\cite{Dickhoff2004, Som2020, PhysRevC.101.014318}. These methods are favored because their polynomial scaling of the computational cost with respect to the system size, and thus have enabled calculations of heavier nuclei. However, they also share some basic limitations. To gain better accuracy, one has to push the truncation schemes to higher orders, making both mathematical expressions and computational cost very challenging. Moreover, for realistic calculations in very large model spaces, providing reliable estimates of many-body uncertainties remains highly challenging~\cite{Hergert2020,y8kt-mgf5}. Although there are many ongoing efforts for higher-order calculations~\cite{PhysRevC.103.044318,PhysRevC.111.034311,PhysRevC.110.044316,PhysRevC.110.044317,PhysRevLett.134.182502,drischler2026,q3vn-8y8s}, the pattern of many-body convergence is still difficult to assess in this scheme.

These limitations motivate stochastic full-space approaches that retain the accuracy of full configuration interaction (FCI) while avoiding explicit storage of the many-body Hamiltonian. In electronic, molecular and various condensed matter systems, the full configuration interaction quantum Monte Carlo (FCIQMC) has been widely applied for more than a decade~\cite{Booth2009,Cleland2010,Cleland2011,PhysRevLett.109.230201,PhysRevB.85.081103,Booth2012,Booth2012n,PhysRevB.90.155130,PhysRevLett.114.033001,Blunt2015,Blunt2015_2,PhysRevLett.118.176403,Blunt2017,PhysRevLett.121.056401,Samanta2018,Guther2020}, and has proven to be one of the most accurate methods in quantum chemistry, especially for strongly correlated systems~\cite{PhysRevX.10.011041}. FCIQMC does not require a fixed-node approximation or a guiding wave function~\cite{Spencer2019}, which is a key difference from some other quantum Monte Carlo (QMC) methods~\cite{RevModPhys.87.1067,PhysRevC.97.044318,PhysRevA.88.053622,PhysRevLett.112.221103,PhysRevC.107.044303}. In FCIQMC, the correct node structure emerges naturally in the simulation, and the fermion sign problem~\cite{Lee2009,Elhatisari2024,pn99-6dxt} is overcome by the walker annihilation algorithm. As an in-principle exact method, FCIQMC retains exponential scaling. Thanks to its nearly perfect parallelization efficiency, FCIQMC calculations can be performed on massively parallel supercomputers, making the study of large systems feasible~\cite{Guther2020}. Preliminary FCIQMC calculation based with-core shell model using phenomenological interactions was encouraging~\cite{Jin2025}, and application to nuclear matter has been successful~\cite{hu2026fciqmc}.

\begin{figure*}[t]
    \centering
    \includegraphics[width=0.95\textwidth]{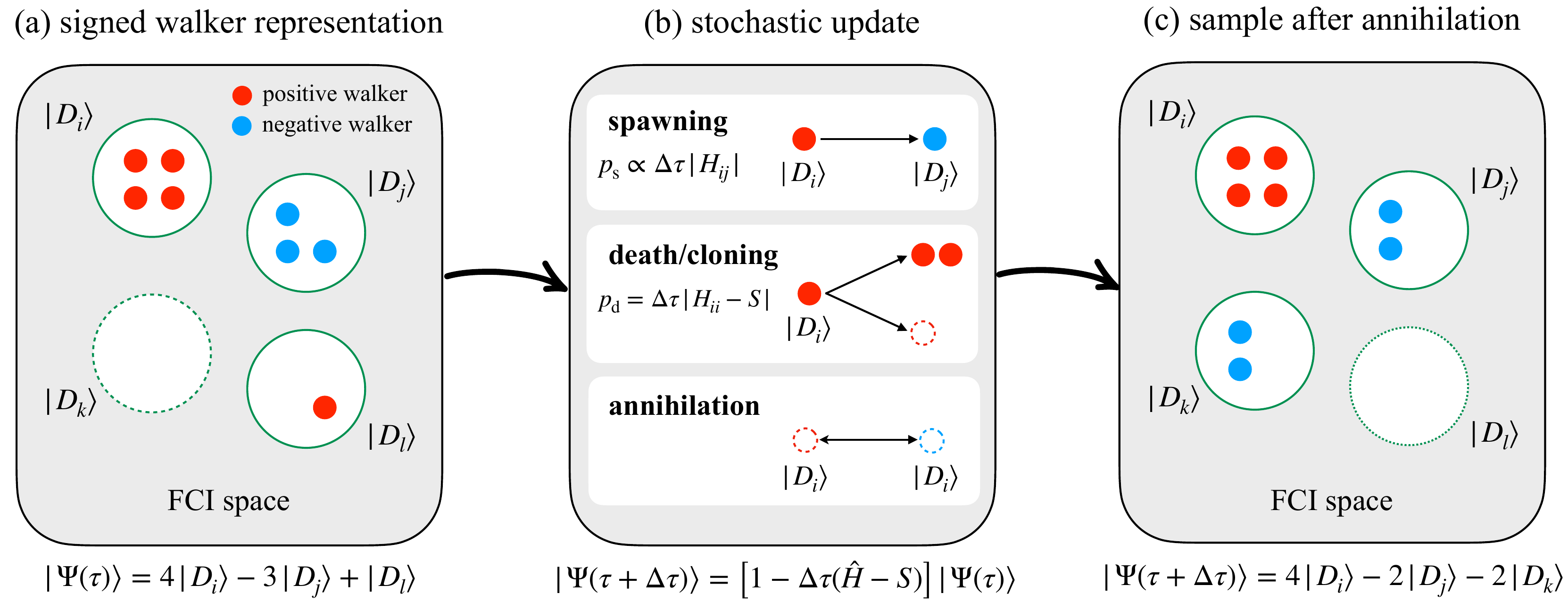}
    \caption{Schematic illustration of the FCIQMC propagation in the FCI determinant space. (a) The wave function is represented by signed walkers on Slater determinants, with the net signed population on $\ket{D_i}$ giving the coefficient $N_i(\tau)$. Red and blue dots denote positive and negative walkers, respectively; empty dashed determinant circles in panels (a) and (c) indicate determinants with zero population. (b) During each imaginary-time step, spawning samples the off-diagonal Hamiltonian couplings, death/cloning samples the diagonal term shifted by $S$, and annihilation cancels walkers of opposite signs on the same determinant. (c) The remaining signed populations after annihilation define the updated stochastic wave function $\ket{\Psi(\tau+\Delta\tau)}$.}
    \label{fig:sketch}
\end{figure*}

In this Letter, we report the first application of FCIQMC to accurate \textit{ab initio} calculations of finite nuclei in strong interactions. Using an optimized chiral two-nucleon interaction at next-to-next-to-leading order~\cite{PhysRevLett.110.192502}, we present accurate ground-state energy and charge radius calculations of $^4$He, $^8$Be, $^{12}$C and $^{16}$O nuclei with sub-percent-level many-body uncertainties estimated. $^{4}$He serves as a classical benchmark nucleus, for which virtually exact calculations can be obtained with the NCSM. $^{8}$Be and $^{12}$C pose significant challenges due to their pronounced clustering structures require the inclusion of very high-order correlations. $^{16}$O is a key nucleus for developing chiral interactions because (i) it is sensitive to partial waves in the interaction beyond those probed in the lightest nuclei ($A\le 4$); (ii) one can build high-fidelity emulators for this nucleus; and (iii) experience has shown that nuclear saturation is difficult to reproduce without this anchor~\cite{PhysRevLett.110.192502,ekstrom2015:sat,jiang2020,Hu2022np,hu2025:texas}. Therefore, accurate benchmarks are essential for assessing the accuracy of chiral interactions developed using polynomial-scaling many-body methods. By systematic comparisons with MBPT, IMSRG and CC methods, we show the crucial role of higher-order correlations in nuclear structure calculations, which are missing in the many-body expansion methods. The benchmarks are particularly valuable for clustering or collective systems in which reference-based many-body expansions may converge slowly. As such, the development of FCIQMC in \textit{ab initio} nuclear theory not only presents an important step towards next-generation many-body methods with higher precision, but also offers a way to advance our understanding of nuclear forces and fundamental symmetries.

\textit{Method.\textemdash}
A detailed description of the FCIQMC algorithm and uncertainty analysis is provided in a companion paper~\cite{companion_prc}. Here we only present a brief summary.

In FCIQMC, we solve the imaginary-time Schr\"odinger equation,
\begin{equation}
-\dfrac{\mathrm{d}}{\mathrm{d}\tau} \ket{\Psi(\tau)} = (\hat{H}-E_0)\ket{\Psi(\tau)},
\end{equation}
to obtain the exact ground-state wave function $\ket{\Psi_0}$ in the long-time limit $\tau\to\infty$. Here $\hat{H}$ is the many-body Hamiltonian, $\tau$ is the imaginary time, and $E_0$ is the unknown ground-state energy. The so-called walkers are introduced~\cite{Booth2009,Booth2012n} to sample the wave function $\ket{\Psi(\tau)}$ dynamically in the FCI basis of all possible Slater determinants (SDs), such that
\begin{equation}
    \ket{\Psi(\tau)} = \sum_{i} N_i(\tau) \ket{D_i},
\end{equation}
where wave function coefficient is represented by $N_i(\tau)$, and the signed sum of those walkers on $\ket{D_i}$. The evolution of the walkers follows the master equation,
\begin{equation}
    N_i(\tau+\Delta\tau) = \big[1-\Delta\tau(H_{ii}-S)\big]N_i(\tau) - \Delta\tau\sum_{j\neq i} H_{ij}N_j(\tau),
\end{equation}
where the unknown $E_0$ has been replaced by a variable $S$, which is known as the shift. Starting from an initial wave function, the walker ensemble is iteratively updated through spawning, death/cloning, and annihilation steps, with a sketch shown in Fig.~\ref{fig:sketch}. Initially, we fix the value of the shift, and the total number of walkers grows exponentially. When the total walker number reaches a desired walker population $N_\mathrm{w}$, we allow $S$ to self-adjust during the simulation, keeping the total walker number roughly constant. After the walker ensemble reaches equilibrium, the evolution still continues until sufficient statistics are accumulated to ensure that statistical uncertainties are sufficiently small. As the ground-state wave function is directly sampled in FCIQMC, any observable can be calculated straightforwardly. In this Letter, we focus on ground-state energy ($E$) and charge radius ($R_\mathrm{ch}$) for $^4$He, $^8$Be, $^{12}$C and $^{16}$O.

In large-model-space calculations, we use the initiator approximation~\cite{Cleland2010} together with the adaptive-shift correction~\cite{Ghanem2019,Ghanem2020} to control the sign problem, which allows controllable and improvable estimates of the FCI limit with the walker number.
By increasing $N_\mathrm{w}$ in the calculation, FCIQMC converges to FCI systematically:
\begin{equation}
\lim_{N_\mathrm{w}\to\infty} \mathrm{FCIQMC}=\mathrm{FCI}.
\end{equation}
Thus, the systematic errors of FCIQMC can be reduced and estimated by increasing this single parameter (see the companion paper for details~\cite{companion_prc}).

\textit{Results and discussion.\textemdash}
An optimized chiral two-nucleon interaction N$^2$LO$_\mathrm{opt}$~\cite{PhysRevLett.110.192502} is used in this work, within a spherical harmonic oscillator (HO) basis defined by $2n+l \equiv e \leq e_\mathrm{max}$ and the HO frequency $\hbar\omega=20$ MeV. There are no configuration truncations in our FCIQMC calculations. We also performed parallel calculations using several other methods for a systematic comparison, including NCSM, second- and third-order MBPT [MBPT(2) and MBPT(3)], CC with singles and doubles (CCSD) and the CCSDT-3 approximation with leading iterative triples corrections~\cite{noga1987}, IMSRG in the normal-ordered two-body operator approximation IMSRG(2) and an approximate IMSRG(3) scheme with perturbative triples included [IMSRG(3f$_2$)+T]~\cite{PhysRevC.110.044317}. The MBPT and CC calculations are only performed for closed-shell nuclei $^4$He, $^{12}$C and $^{16}$O. For IMSRG, the single-reference formulation is used for $^4$He and $^{16}$O, and the valence-space formulation~\cite{PhysRevC.85.061304} is used for $^8$Be and $^{12}$C.

\begin{figure}[t]
\centering
\resizebox{0.48\textwidth}{!}{
\includegraphics{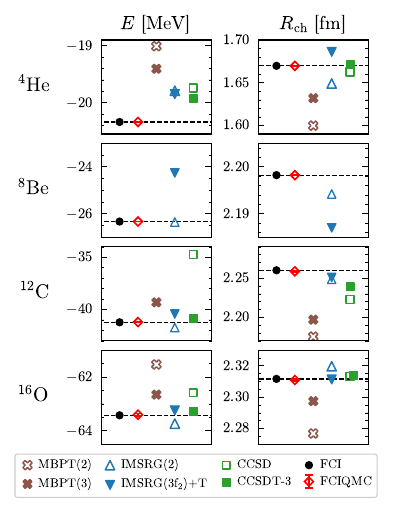}}
\caption{Ground-state energies ($E$) and charge radii ($R_\mathrm{ch}$) for $^4$He, $^8$Be, $^{12}$C and $^{16}$O nuclei computed using the interaction N$^2$LO$_\mathrm{opt}$ in the small $e_\mathrm{max}=2$ model space. FCIQMC results are  compared with MBPT(2), MBPT(3), IMSRG(2), IMSRG(3f$_2$)+T, CCSD, CCSDT-3 and FCI. The FCIQMC error bars denote the estimated many-body uncertainties and are too small to be visible. Dashed horizontal lines represent FCI results used to guide eyes.}
\label{fig:comparison_emax2}
\end{figure}

\begin{figure}[t]
\centering
\resizebox{0.48\textwidth}{!}{
\includegraphics{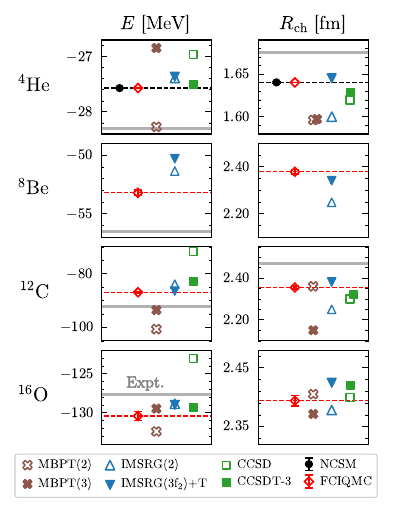}}
\caption{Similar to Fig.~\ref{fig:comparison_emax2}, but for the large $e_\mathrm{max}=10$ model space. FCI results are not available in this model space, and converged NCSM results are available only for $^4$He. The black, red, and gray horizontal lines represent NCSM results, FCIQMC results and experimental data~\cite{Wang_2021}, respectively.}
\label{fig:comparison_emax10}
\end{figure}

Figure~\ref{fig:comparison_emax2} shows the results obtained in a small $e_\mathrm{max}=2$ model space. In this case, FCI results can be obtained by direct diagonalization. For all ground-state energies and charge radii, FCIQMC always gives accurate results in agreement with FCI, indicating that the FCI wave functions are accurately sampled. Among all adopted many-body expansion methods, MBPT generally shows the largest deviation. IMSRG results are generally reasonable, with an exception of $^8$Be where higher-order IMSRG(3f$_2$)+T results deviate even more strongly from the exact values. For CC, a systematic improvement can be seen when changing from CCSD to CCSDT-3. However, for the ground-state energy of $^4$He and the charge radius of $^{12}$C, the many-body convergence of CC still seems slow.

We now turn to large-model-space calculations with $e_\mathrm{max}=10$, which is sufficient to achieve single-particle basis convergence in the present calculations. About $10^{6}$ CPU hours were used to perform these FCIQMC calculations on advanced supercomputers. In this case, FCI results are unavailable because the Hilbert space is prohibitively large: for $^{16}$O alone, the FCI dimension is $9.7\times 10^{32}$, far beyond explicit diagonalization. As shown in Fig.~\ref{fig:comparison_emax10}, our FCIQMC results for $^4$He agree perfectly with virtually-exact $N_\mathrm{max}$-extrapolated NCSM results. For heavier nuclei, NCSM fails to converge, whereas FCIQMC still provides accurate results with sub-percent estimated many-body uncertainties. For other many-body expansion methods, the comparison is similar to that in the small-model-space calculations. MBPT and IMSRG show no clear convergence when going to higher orders, and there is a systematic improvement for CC when changing from CCSD to CCSDT-3. However, for the charge radius of $^{16}$O, the inclusion of leading iterative triples corrections in CC leads to an even larger deviation compared with our FCIQMC result, as also indicated by the small-model-space comparison.

The comparison for $^8$Be is particularly interesting because its ground state has a pronounced molecular structure~\cite{PhysRevLett.131.212501,PhysRevLett.134.162503}. The IMSRG(2) underestimates its ground-state energy, and when going to IMSRG(3f$_2$)+T the deviation is even larger. This indicates that omitted higher-order collective correlations are important for describing clustered nuclear states. For IMSRG, such correlations may require deformed or multi-reference formulations~\cite{Frosini2022,PhysRevC.105.064311,PhysRevC.105.L061303,sun2025} and/or the inclusion of higher-body operators beyond the present truncation. However FCIQMC includes many-body correlations to all excitation ranks within the chosen basis, which allows it to remain accurate when treating collective states.

The above systematic benchmarks and comparisons not only show the accuracy and power of the FCIQMC method, but also provide valuable information for other many-body methods. In previous \textit{ab initio} nuclear structure calculations, it is almost impossible to provide a reliable estimation of the many-body uncertainty, except for very light nuclei. By comparing different methods, it has been roughly estimated that the uncertainties of state-of-the-art many-body expansion methods are on the order of 1--2\% for ground-state energies and 1--1.5\% for charge radii, when using low-momentum interactions, consistent with the estimations in Refs.~\cite{Hergert2020,PhysRevC.111.034311}. For collective or deformed states, the convergence of many-body expansions is even slower. In this sense, FCIQMC not only presents a next-generation \textit{ab initio} many-body method with higher accuracy, but also shows the crucial role of higher-order correlations, and provides exact benchmarks for the further development of other many-body methods.

\textit{Summary and outlook.\textemdash}
In this Letter, we introduce the FCIQMC method to \textit{ab initio} nuclear structure calculations of strongly interacting systems for the first time. As a stochastic projector QMC method in the full configuration space, FCIQMC directly samples the FCI wave function without a fixed-node approximation or a guiding wave function. We have performed calculations for ground-state energies and charge radii of $^4$He, $^8$Be, $^{12}$C and $^{16}$O nuclei on massively parallel supercomputers, obtaining accurate results with sub-percent many-body uncertainties. For other many-body expansion methods, including MBPT, IMSRG and CC, the many-body errors due to their truncation schemes are directly revealed through comparison with FCIQMC. The $^8$Be results further demonstrate the advantage of FCIQMC for clustering states in which high-order correlations slow down the convergence of reference-based expansions.

As a precise \textit{ab initio} many-body method with quantified many-body uncertainties, FCIQMC will have a broad impact on nuclear structure theory. It connects the microscopic nuclear forces with many-body observables directly, shedding light on more reliable evaluation and future development of nuclear forces. FCIQMC may also be used to solve some long-standing puzzles in nuclear theory, such as the evolution of charge radii in the calcium chain~\cite{GarciaRuiz2016} and the calculation of isospin-breaking corrections in superallowed Fermi decays~\cite{PhysRevC.66.024314,Stroberg2021,PhysRevC.102.045501}, where the central challenge is high-order many-body effects.

\textit{Acknowledgments.\textemdash}
We thank S. L. Jin, J. H. Hou, Z. Y. Meng, J. G. Li, J. D. Holt and F. Marino for useful discussions. We thank Gaute Hagen for providing us with the coupled-cluster code for CCSD and CCSDT-3 calculations. 
The NCSM, MBPT and IMSRG calculations were performed using the \texttt{Bigstick}~\cite{code-bigstick}, \texttt{HartreeFock}~\cite{code-mbpt} and \texttt{imsrg++}~\cite{code-imsrg} codes, respectively.
This work was supported by the National Key R\&D Program of China under Grants Nos. 2024YFA1610900 and 2023YFA1606401; the National Natural Science Foundation of China under Grants Nos. 12335007 and 12535008; the Fundamental Research Funds for the Central Universities, Peking University. Numerical calculations in this work were performed at the Huabei Advanced Computing Center and the High-Performance Computing Platform of Peking University.

\textit{Data availability.\textemdash}
The data that support the findings of this article are openly available~\cite{data}.

\bibliographystyle{modified-apsrev4-2.bst}
\bibliography{reference}

\end{document}